# DeepLink: A Novel Link Prediction Framework based on Deep Learning

Mohammad Mehdi Keikha[1,2], Maseud Rahgozar[1], Masoud Asadpour[1]

Email: {mehdi.keikha, rahgozar, asadpour}@ut.ac.ir

Corresponding Author: Maseud Rahgozar

**Abstract**:

Recently, link prediction has attracted more attentions from various disciplines such as computer science, bioinformatics and economics. In this problem, unknown links between nodes are discovered based on numerous information such as network topology, profile information and user generated contents. Most of the previous researchers have focused on the structural features of the networks. While the recent researches indicate that contextual information can change the network topology. Although, there are number of valuable researches which combine structural and content information, but they face with the scalability issue due to feature engineering. Because, majority of the extracted features are obtained by a supervised or semi supervised algorithm. Moreover, the existing features are not general enough to indicate good performance on different networks with heterogeneous structures. Besides, most of the previous researches are presented for undirected and unweighted networks.

In this paper, a novel link prediction framework called "DeepLink" is presented based on deep learning techniques. In contrast to the previous researches which fail to automatically extract best features for the link prediction, deep learning reduces the manual feature engineering. In this framework, both the structural and content information of the nodes are employed. The framework can use different structural feature vectors, which are prepared by various link prediction methods. It considers all proximity orders that are presented in a network during the structural feature learning. We have evaluated the performance of DeepLink on two real social network datasets including Telegram and irBlogs. On both datasets, the proposed framework outperforms several structural and hybrid approaches for link prediction problem.

## 1. Introduction:

Online Social networks (OSNs) are one of the fastest growing industries around the world. OSNs such as Facebook, Twitter and Telegram provide diverse services to exchange information. Because of dynamic nature of OSNs, their mining and analyzing have attracted more attentions. In recent years, many researchers from different disciplines have conducted many experiments on OSNs to extract valuable knowledge from them. Link prediction is one of the interesting analysis tasks on OSNs that finds unknown links on a social network based on the existing links and nodes' attributes [1]. It discloses network topology evolution over the time and discovers unknown links in the current state of the network. Link prediction is used for diverse applications like biological networks [2], [3], friend suggestion[4], [5], recommendation systems [6], [7].

---

[1] Database Research Group, Control and Intelligent Processing Center of Excellence, School of Electrical and Computer Engineering, College of Engineering, University of Tehran, Tehran, Iran

[2] University of Sistan and Baluchestan, Zahedan, Iran

Many supervised and unsupervised methods have been proposed for the link prediction problem. Existing link prediction methods can be classified in similarity based methods and learning based algorithms. In similarity based methods, a similarity function is defined to measure the probability of link existence between any pair of nodes based on numerous information such as structural and content information. While the learning based methods extract various features to build a model for the given network. Then, unknown links would be predicted by the learned network model.

A considerable part of previous researches employ the structural properties of the given network to measure the similarity of nodes. A notable number of the topology based methods only consider the local structure such as common neighbors (CN) of two nodes as the measure of similarity [1], [8]–[10]. Some researchers employ comprehensive structural information of the networks, including paths [11], [12] and communities [13] as a basis to find the similarity of nodes. The time complexity of the global structural based methods is higher than CN based methods. Rebaza et al. [14] integrated community information with local structure for link prediction. The structural based methods entirely rely on the domain and topology of the given network. Thus, these methods show different performance for each network because of the structural information loss in the given network.

In [15], the effects of non-structural information are investigated for the link prediction. They found that contextual information dissemination can change the structure of network and vice versa. A number of researchers utilize profile information and user generated contents like gender, organization, tweets and blog posts as contextual information to measure the similarity of two nodes [16]–[18]. After the similarity computation between all the nodes' pairs, the highest similarity scores are selected as missing/future links. These algorithms do not leverage structural information for the link prediction simultaneously.

In the recent years, hybrid link prediction methods [15], [19] as learning based approaches are proposed which incorporate both the structural and content information. Feature engineering is the key component of these methods. While the accuracy of these algorithms is better than the previous researches, but the process of feature extraction is supervised and time consuming. Thus, they are not scalable for large scale networks. Furthermore, the process of how to extract and combine the structural and non-structural information is very important. It is notable that more features result in better prediction of unknown links, but extraction and selection of them lead to higher time complexity. Moreover, most of the previous researches disregard to weights and directions of the edges on the network. Whereas in real social networks, a person based on the degree of his/her interests may follow others who are not familiar with each other to some extent. Thus, there is a weighted and directed link between them.

Concerning with the above challenges, in this paper, we propose a framework called "DeepLink" to predict missing/future links by using deep learning techniques. In recent years, deep learning techniques have changed the performance of different applications such as speech recognition [20], image processing [21], information retrieval [22] and social network analysis [23]. One of the most interesting applications of deep learning is network embedding, which encodes the local and global structural information of the network into feature vectors [23], [24]. The learned feature vectors are employed in different applications such as clustering [25] and link prediction [23], [26].

To learn the best feature vector for content information of the nodes, Doc2Vec algorithm is used [27]. The combination of structural and content feature vectors is used to generate a unified feature vector for each node. Finally, a classifier is learned by the integrated feature vectors to predict unknown links of the given network.

We have evaluated DeepLink framework on Telegram and irBlogs datasets. The empirical analysis indicates significant improvements on both datasets in comparison to the previous researches on the link prediction.

To summarize, we make the following contributions:

- We present an unsupervised link prediction framework called "DeepLink" that extracts the best feature vectors from different types of networks such as weighted, directed and complex networks. DeepLink is also scalable for large networks.
- DeepLink is a general link prediction framework that employs both the structural and content information of the given network. While most of the previous researches consider some local structural information; we utilize community and neighborhood of nodes as global and local structural information alongside the contents which are generated by the users.
- To the best of our knowledge, DeepLink is the first framework that utilizes deep learning techniques to extract best structural and content features in the link prediction problem.
- The proposed framework is also able to embed the feature vectors that are prepared by the previous researches to predict unknown links.
- We empirically evaluate DeepLink on two different real-world social networks. The experimental results verify the efficiency of the proposed framework in contrast to the other link prediction approaches.

The rest of paper is organized as follows: In section 2, we summarize related works to link prediction methods and network representation learning techniques. Section 3 presents a formal definition of link prediction problem. We explain the details of DeepLink in section 4. Section 5 outlines the experimental results of the proposed framework on two real social networks. Finally, Section 6 presents conclusion and future works.

## 2. Related works:

As previously stated, link prediction methods define a similarity function for a pair of nodes. A substantial number of link prediction methods rely entirely on local structural information. [1] proposed several measures based on structural information of nodes. One of the most frequently used measures for the link prediction problem is common neighbors (CN) which is proposed by [8]. In CN, the number of shared neighbors between two nodes are considered as the measure of similarity. There are many normalization schemes for CN to enhance the accuracy of link prediction such as Jaccard coefficient and Sorensen's index, which are described by [28]. In [8], Preferential attachment [29] and Resource allocation [30] measures, a weight is assigned to each shared neighbor to calculate the probability of a link between two nodes. [31] recently have proposed a novel similarity measure based on the tree augmented naive (TAN) Bayes probabilistic model. It makes better link predictions since the model considers the relation among shared common neighbors to some extent.

Number of researches considers global structural information such as paths and random walks between two nodes as the score of nodes' similarity. [32] proposed local path method that uses two and three hop neighbors of network nodes to obtain the similarity of each pair of nodes [1] considered all different length paths between nodes as the similarity score. [33] proposed SimRank method, which supposes that two nodes are similar if they are connected to the same nodes. In their method, two random surfers are used, which are started from source and destination nodes to measure how soon the random surfers are expected to visit the same node.

Community information is another global structural information, which is employed as useful additional information in many researches. [14] used both follower/followee relations and community information to predict unknown links in twitter network. They use the label propagation algorithm [34] to detect different communities of large networks. In their method, a measure based on Bayesian theorem is proposed to leverage inter and intra community links to predict missing/future links on the network. They found that within community links have more contributions on link prediction than inter community links. While the proposed method considers all the structural information, but the content information is ignored. Thus, their method depends on the network structure, and it would not be appropriate for different network structures. Though, the global structural based methods are more accurate than local neighbors measures due to including more information, but they have scalability issue because of their time complexity [28].

Recently, hybrid methods are proposed for link prediction, which train a classifier by incorporating different features such as topology and contextual information of nodes [35], [36]. Featuring engineering is the key component of hybrid link prediction methods.

[19] recently have proposed a number of weighted measures that employ topological information with the combination of contextual and temporal information on the given network. They first compute the probability of link's existence between nodes based on profile and gender information as contextual information, time attribute of the current links as temporal information and common neighbors of nodes as topological features. Then, a weighted combination of these scores are used to predict links. The proposed method only considers local structural information while as it is stated, community information would enhance the accuracy of link prediction methods. Besides, user generated contents are also ignored in the proposed method.

[15] proposed a hybrid link prediction method that integrates content and structural information. They define a user to user topic inclusion degree (TID) score, which measures the relatedness of two users based on user generated contents. Then, a new network is constructed based on TID scores of the network users. Finally, matrix factorization techniques are employed to fuse adjacency matrix with TID matrix and predict missing/future links based on the learned model from the unified matrix of structural and content information. The proposed method does not consider global structural information. Moreover, the process of TID network construction is time consuming and the proposed method would not be employed for link prediction on real large scale social networks.

In recent years, deep learning techniques are widely used for feature engineering and dimensionality reduction in social networks [23], [24], [26], [37], [38]. In these methods, a number of predefined random walks are generated for each node by different graph search strategies. Then, the walks are used as contextual information in the skip-gram model [39] to learn final structural feature vector of nodes. The final structural feature vectors are employed to detect missing and future links of the network.

Skip-gram model was first used in DeepWalk algorithm for network embedding [23]. In DeepWalk, DFS like search strategy is used to generate random walks. However, the global network structure is not preserved because the community information of network nodes is not considered during the path generation. [37] proposed LINE algorithm, which employs 1 and 2 hop neighbors to learn nodes' feature vector, but they also preserved local information of the networks. In DeepWalk and LINE, the global structure of networks such as social theories properties are ignored and only the local information of nodes is employed to learn the best feature vector. [26] in Node2Vec generated random walks based on DFS and BFS like strategies. They also consider first and second order proximities during search strategies.

In M-NMF method [38], modularized non negative matrix factorization is applied to preserve both local neighborhood and community structure of the network. Wang et al. learn local and community structure based on two optimization function separately. Consequently, they combine the final representations. Their final representation is not general enough to be used in different network analysis tasks. M-NMF also has some local structure information loss because it combines first and second order proximities in a unified matrix. Unification of matrices leads to information loss for different proximities during representation learning. Their method also suffers from scalability when the networks are large because they should learn many parameters to preserve local and global structures.

Overall, many researches are conducted on the link prediction problem. However, limited number of researches employ all the information, which are embedded into the network simultaneously. Most of the previous researches for link prediction are not considered either the global structure of the network or the user generated contents. Besides, most of the previous researches which regard to structural and content information concurrently, face with scalability issue to predict unknown links in large networks. Moreover, choosing best features on different networks is another challenging task, which is not properly investigated by the previous researches. In this paper, an unsupervised link prediction framework is presented to extract both local and global properties of the network structure along with the contents that each user shares on the network. In DeepLink framework, the best features for each node are extracted by using deep learning methods. Before DeepLink framework is introduced in section 4, the formal definition of link prediction is presented in section 3.

## 3. Problem Definition:

Suppose we have a network graph $G = (V, E)$ in which each edge $e = (u, v, w) \in E$ represents a collaboration between $u$ and $v$ with weight $w$. In real networks, different attributes including time stamps, probability of message forwarding and the number of message passing between users are considered as weight $w$. However, a number of edges may not have any edge attribute. In such condition, $w$ is equal 1 for all the network edges.

In addition to the links in the network, each user $u$ has other features such as user-generated contents and profile information. In order to consider those features, a feature vector $A$ is defined for each node, which hold contextual properties, including contents and profile information about the node. The probability of links between nodes can be calculated based on the similarity of their associated feature vectors. Two snapshots of the network at time $t_1$ and $t_2$, on the condition that $t_1 < t_2$, are considered to extract the data set $D = \{x_1, x_2, \ldots, x_m\}$ and its corresponding label set $Y = \{y_1, y_2, \ldots, y_m\}$. $x_i$ is the feature vector of any node pair $(u.v)$ who has no collaboration links at $t_1$ and $y_i \in \{0, 1\}$ is the label of $x_i$. If $u$ and $v$ form a collaboration link at $t_2$, the value of $y_i = 1$, otherwise $y_i = 0$. Thus, the link prediction is the process of learning a classification model using data set $D$, then the model outputs a list of edges not presented at the time $t_1$, While they are in the network at the time $t_2$.

## 4. DeepLink framework:

In this section, the architecture of our proposed link prediction framework will be presented. To extract the best features for the link prediction, both the structural and content information of network is used. In order to learn the best structural features, different levels of structural information are employed. In our framework, a number of customized paths are generated to capture n-hop neighbors and community information for all the nodes in the network by using an extended version of CARE algorithm for the network embedding [24]. In contrast to previous researches in network representation learning, edge weights are also considered during the path generation and feature learning in DeepLink. Afterward, the best textual features are learned from the user-generated contents by using Doc2Vec algorithm [27]. The structural and content feature vectors are combined to obtain the best nodes' feature vector. The final vectors are used to learn a classifier to predict missing/future links in the given network. The architecture of DeepLink framework for the link prediction is illustrated in figure 1.

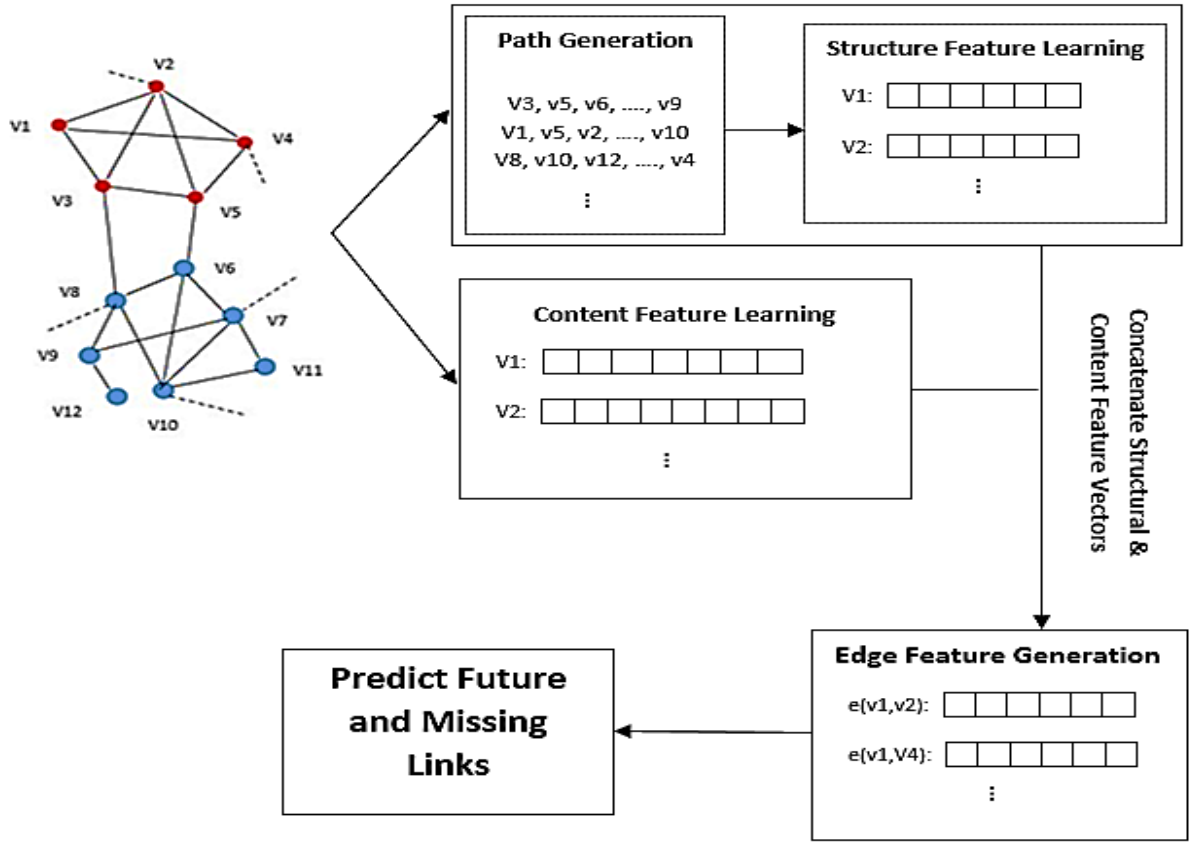

*Figure 1- Architecture of DeepLink*

In the following subsections, we will explain each component of DeepLink in details.

### 4.1. Structural feature learning:

As mentioned in section 2, the structural features, including local and global information are widely used in link prediction methods. In DeepLink, a new network embedding method is used to capture structural information during feature learning [24]. To extract structural feature vectors, the belonging communities of each node are first detected by using Louvain algorithm [40]. Afterwards, the associated communities of each node along with the local neighborhood are used to generate a number of custom paths. To extract the structural feature vectors, Word2Vec deep learning model [39], [41] is used by DeepLink.

In the other words, a network is considered as a text document which its nodes are vocabulary of the document and custom paths thought out as sentences. A custom path simulates a random walk which involves n-hop neighbors of network nodes as well as the nodes that belong to the same community with the current node on the path. In Louvain community detection method, graph clusters are detected by modularity maximization method. The main assumption of the modularity maximization algorithm is that the nodes of the same cluster have more links with each other in comparison to the nodes of other clusters. On this basis, the nodes on a custom path have strong relationship with each other based on local structure and homophily properties. Finally, we generate a predefined number of custom paths for each node to learn the structural feature vectors by Word2Vec model.

The process of customized path generation in DeepLink is shown in figure 2.

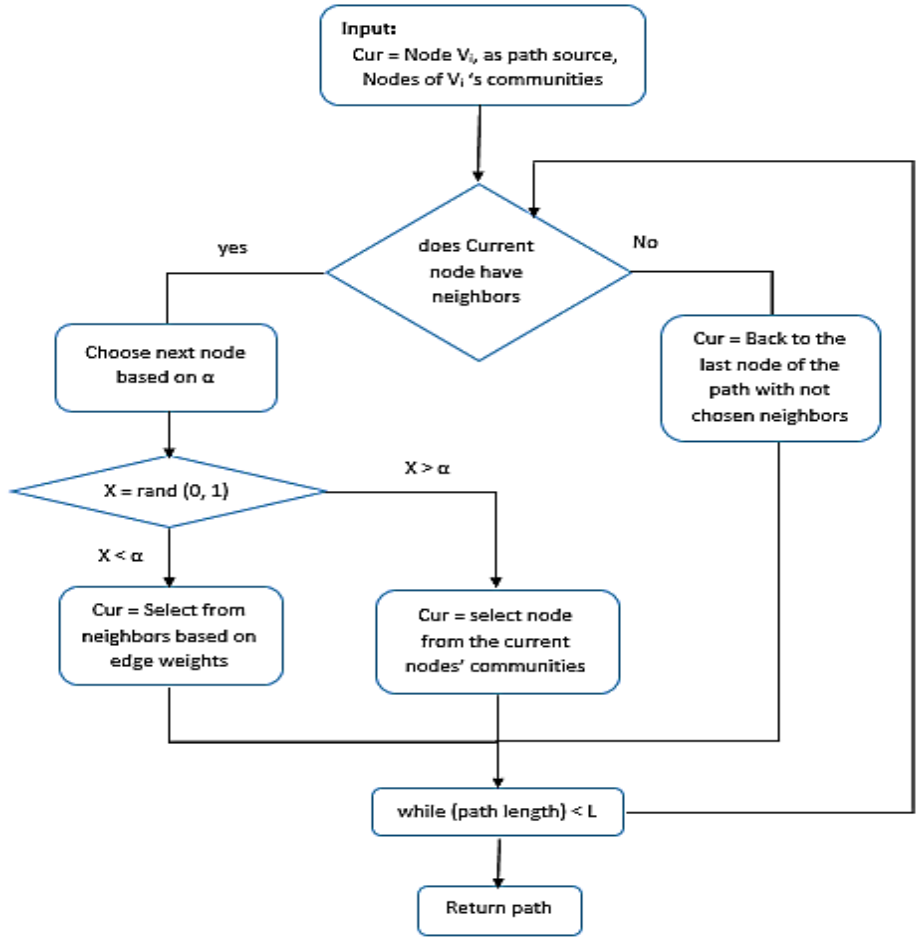

*Figure 2- Custom path generation for structural feature learning*

In figure 2, the next node on the path is chosen based on the threshold value of α. If $X \leq α$, the next node is selected from the direct neighbors of the current node on the path. Otherwise, a node from the communities of the last node on the path would be selected. In DeepLink, the length of a custom path is restricted to variable L. This means that a custom path with the length of L is able to consider n-hop neighbors along with the nodes of the belonging communities.

During the path generation, we consider the edge weights as the probability of visiting nodes. While the previous network representation learning researches ignore this valuable information. Edge weights indicate the probability of interacting two nodes. So, node selection based on the edge weights will enhance the quality of the generated paths because the paths contain the nodes that have strong ties with each other than the other network nodes. Since the paths are considered as the sentences in Word2Vec model, more meaningful paths as inputs of Word2Vec algorithm result in better structural feature vectors.

When the specified number of paths are generated for each node, Word2Vec algorithm is used to learn the structural feature vectors [39]. In Word2Vec algorithm, two nodes are similar when they are visited in many paths with each other. The probability of the current nodes' neighbors in the generated path is maximized using Eq. 1:

$$P(N(u) | f(u)) = \max_{f} \prod_{\substack{j=i-w \\ j \neq i}}^{i+w} P(v_j | f(u)) \quad N(u) = \{v_{i-w}, \ldots, v_{i+w}\} \setminus u \quad (1)$$

Where $w$ is the length of a context window which slides over the custom paths. The best features of node $u$ are kept in $f(u)$. In addition to, $P(v_j | f(u))$ is the conditional probability of visiting the node $v_j$ given the feature vector $f(u)$. In Eq. 1, the edge weights are employed during feature learning based on the following formula:

$$P(v_j | f(u)) = 1 / (1 + e^{-f(u).f(v_j) \cdot w(u.v_j)}) \quad (2)$$

According to Eq. 2, we consider the edge weights for a pair of nodes that have direct links on the network. For the nodes that belong to the same community, the edges' weights are considered 1. In Eq. 2, $w(u.v_j)$ is the weight of edge $(u.v_j)$ which is calculated using Eq. 3:

$$w(u.v_j) = \begin{cases} w(u.v_j) & if \ e(u.v_j) \in E(G) \\ 1 & otherwise \end{cases} \quad (3)$$

When the structural feature vectors are learned in DeepLink, the content feature vectors are extracted by DeepLink. In the following, the content feature engineering is explained.

### 4.2. Content feature learning:

In this section, detail of our content feature learning module in DeepLink framework is explained. We first integrate all the contents, which are shared by a user into one document. In the generated document, each post of a user with its different attributes such as timestamp thought out as a paragraph. Then, Doc2Vec algorithm is used to learn the best content feature vectors of network nodes [27]. The learned features depict the topics and interests of the user on the network. So, two users may link to each other based on their interests.

One of the main contributions of DeepLink is the usage of deep learning techniques to learn the latent feature vectors. In the latest researches [15], LDA algorithm is used to indicate the topics' similarity of two users on the network. Whereas LDA illustrates poor performance in comparison to deep learning techniques on topic modeling. Because LDA uses frequency of the words in the document to extract the topics, but the deep learning based methods use the semantic of each word on the given context which leads to better topic modeling on the network.

Doc2Vec extends Word2Vec algorithm for larger text blocks such as sentences, paragraphs, and documents. In content feature learning, a feature vector is generated for each post of user, then the learned vectors are employed to learn the next word's feature vector in the given sentence. Figure 3 indicates how the post's feature vector and words' feature vector are contributed to learn the feature vector of the next word in the sentence.

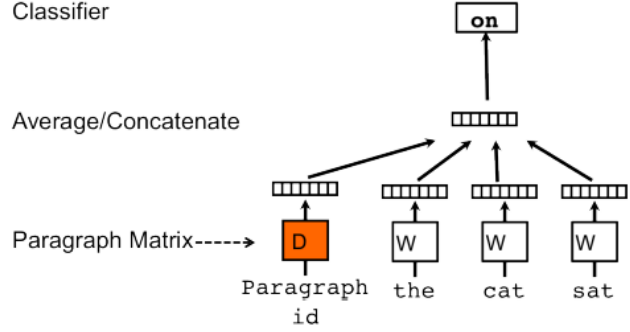

*Figure 3- Content feature vector learning in DeepLink* [27]

In this figure, contextual information is embedded through the paragraph matrix during the learning of the next word's feature vector in a sentence. Suppose that there are $n$ words in the vocabulary. The feature vector of the next word in the sentence is predicted by maximizing the average log probability based on Eq. 4:

$$\max \frac{1}{n} \sum_{i=k}^{n-k} \log p(w_i \mid w_{i-k}, \dots, w_{i+k}) \qquad (4)$$

Where $w_i$ is a word in a context window with the length of $k$ words. The occurrence probability of the word in the context window is computed by Eq. 5:

$$p(w_i \mid w_{i-k}, \dots, w_{i+k}) = \frac{e^{y_{w_i}}}{\sum_i y_i} \qquad (5)$$

Where $y_i$ is the feature vector which is generated by concatenation of the words' feature vector and the paragraph feature vector in Eq. 6:

$$y_i = b + Uh(w_{i-k}, \dots, w_{i+k} . d_i\, ; W.D) \qquad (6)$$

In Eq. 6, $W$ is a feature matrix in which each column holds the feature vector of a word in the vocabulary; While $D$ is the feature matrix of the posts which are generated by a user on the network. We extract and learn the text feature vector that reflects the topics of interests for each node.

After learning the structural and content features for each node, two feature vectors are concatenated to form the final feature vector of nodes. The learned feature vector encompasses various information about the node, including structural and content properties. It is worth noting that the feature vectors are learned for each user. While, in the link prediction, we should learn the edge feature vectors. To obtain the edge feature vectors, the Hadamard operator is employed based on Eq. 7.

$$Hadamard: \quad [f(u) \boxdot f(v)]_i = f(u)_i * f(v)_i \qquad (7)$$

The Hadamard operator is the best operator for edge feature learning from user feature vectors [24], [26]. An edge feature vector indicates that the source and destination nodes of a link are similar based on their structural and content features.

## 5. Experimental results:

In this section, DeepLink framework experimental results on two real social network dataset are compared with the state-of-the-art methods for link prediction, which are introduced in section 5.1. To evaluate the framework, two different datasets including Telegram and irBlogs are employed, which are introduced in subsection 5.1. We also provide detailed discussions about the framework in section 5.2. Finally, the parameter sensitivity of DeepLink is investigated in section 5.3.

### 5.1. Baseline algorithms:

In the experiments, the structural and content feature vectors are learned independently, then the learned feature vectors are combined to form the final feature vector of nodes. To do fair evaluation, we first learn the structural feature vectors by the following network embedding methods. Then, the learned content feature vectors are concatenated to the structural feature vectors to predict unknown links by different hybrid algorithms.

**Node2Vec** [26]: Grover and Leskovec proposed to learn structural feature vectors by generating second order random walk. Node2Vec ignores the community information and higher order proximities during path generation.

**DeepWalk** [23]: DeepWalk is the first algorithm for network embedding, which uses Word2Vec model to learn structural feature vectors. In DeepWalk, the communities of network are disregarded during the path generation.

**LINE** [37]: In LINE algorithm, nodes' feature vectors are generated by optimizing two independent functions for the first and second order proximities. Next, combinations of two functions are employed to provide final structural feature vectors. LINE is also neglect to community information of network topology.

**M-NMF** [38]: Wang et al. proposed a method that considers community structure of the given network alongside the first order proximities in the network representation learning process. The feature vectors of the community structure are combined with the nodes' feature vectors to form the final network embedding. In M-NMF method, non-negative matrix factorization is used to extract final feature vector.

As it is previously stated, in DeepLink framework, the learned structural feature vectors can be obtained by different algorithms for the link prediction problem. So, we compare the performance of structural feature vectors, which are prepared by DeepLink, against to the frequently used local structure based methods for the link prediction, including Common neighbors, Jaccard, Preferential attachment, Adamic-Adar, and Sorensen's index. The local structural based methods compute a score for each pair of nodes in the tests sets. Then we sort the node pairs based on their scores in a decreasing order.

### 5.2. Datasets

In the experiments, we adopt two unexplored social network datasets, i.e. irBlogs [42] and Telegram. Each of them is directed and contains information on two aspects; the network topology and users generated contents.

**Telegram[3] dataset**: Telegram is a cloud-based social network that users can send messages and exchange photos, videos, stickers, audio and files of any type. Channels are one of the interesting features of Telegram, which can be created for broadcasting messages to an unlimited number of subscribers. Each message in a channel has its own view counter, showing how many users have seen this message [43]. We have crawled and preprocessed two snapshots of channels' forward messaging communications for two different months. Then, predict the links which are made during two snapshots. This dataset is a directed and weighted network that shows the relation of different public channels in Telegram. Each node is a Telegram channel and the edge $(u.v.w)$ indicates the node $u$ shares $w$ messages which are generated by node $v$.

---
[3] Telegram.org

In this dataset, the posts of channels are preprocessed and concatenated to extract content feature vector. To generate the structural feature vector, we consider all the links, which are made in the new snapshot of the dataset as the positive test set. Some non-existing links in the two snapshots are also chosen as the negative test set with the same size of the positive test set. The training set involves all the existing edges in the first snapshot of the network as the positive train set and non-existing edges as the negative train set with the same size of the positive train set. In the experiments, the content feature vector of the Telegram dataset is combined with different structural feature vectors, which are generated by the baseline algorithms.

**IrBlogs dataset** [42]: It contains nearly 5 million posts and the network of more than 560,000 Persian bloggers. In this dataset, each node is a blog and the edge $(u.v)$ shows that node $u$ shares $w$ contents which are generated by node $v$. For the link prediction, we have randomly removed a portion of existing links as the positive test set and consider all the remained edges in the network as the positive train set. We create a negative train set with the randomly selected nodes which are not existed in the positive train and test sets. Finally, the negative test set is chosen from the edges that are not present in each of positive and negative train and test sets. Because of sparsity issue of the original network, we use a subgraph of irBlogs which has enough edges. Table 1 summarizes the statistics of the datasets which are used for our evaluations.

*Table 1- the statistics of datasets*

|  | |V| | |E| | Directed | weighted |
|---|---|---|---|---|
| Telegram | 93268 | 156851 | ✓ | ✓ |
| irBlogs | 3036 | 11348 | ✓ | ✗ |

To evaluate the accuracy of our proposed method with others, we used the AUC score which can be interpreted as the probability that a randomly chosen missing link is given a higher similarity score than a randomly chosen pair of unconnected links [44]. AUC score is calculated as:

$$\frac{n' + 0.5\, n''}{n} \quad (7)$$

Where $n$ is the number of times that a pair of links from missing/future links set and non-existing links set are chosen randomly. $n'$ is the number of times that the missing/future link got a higher score than the non-existing link and finally, $n''$ is the number of times that they are equal scores.

### 5.3. Parameter settings

To compare our results with the baseline feature learning methods, we have used the same parameter settings that are used in [45], i.e. We choose context window $w=10$, number of generated paths for each node $\mu=10$ and length of the custom path l=80. The optimal value for $\alpha$ is 0.2. In Node2vec algorithm, the best values for p and q are chosen from {0:25, 0:5, 1, 2, 4} as stated in [26]. It is worth noting that the optimal size of structural and content feature vectors is chosen d=100, which would explain in section 5.5. Finally, we use binary regression classifier to predict unknown links.

### 5.4. Experiment results and analysis:

Figure 4 shows the AUC score of different methods in the proposed link prediction framework on Telegram dataset. In figure 4, different structural feature vectors which are obtained by the baseline network representation learning algorithms, are concatenated with the same content feature vector. The final results are indicated in figure 4.

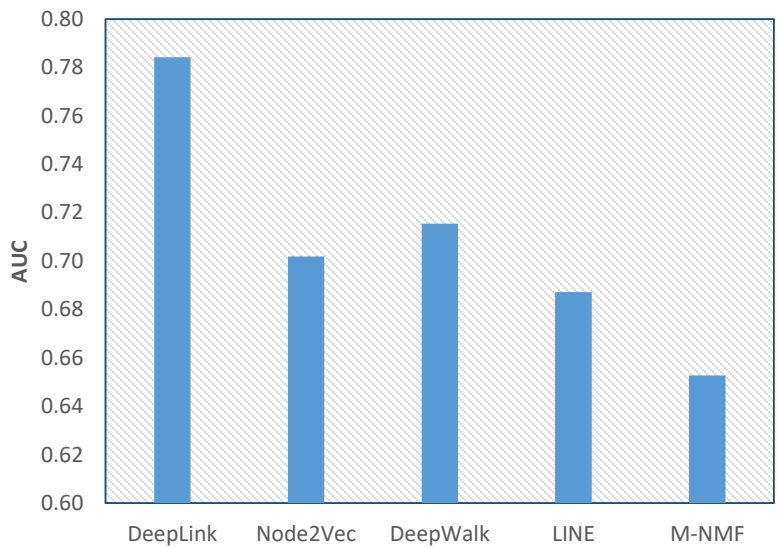

*Figure 4- AUC score for different network representation learning algorithms on Telegram*

As it is shown in figure 4, DeepLink achieves better performance in contrast to the other methods because both local and global structure of the network are employed in DeepLink framework. DeepWalk has the better performance in comparison to Node2Vec. Since different proximity orders of a node are used in DeepWalk, while in Node2vec, first and second order proximities are only considered for feature extraction. Though more general information of the graph is captured by DeepWalk, but the community structure is ignored during feature learning. Even though LINE performs better than M-NMF because of using first and second order proximities, nevertheless, the community structure of the graph is ignored. Additionally, in LINE algorithm, the links that connect two network nodes with the shortest distance of more than two cannot be recognized because the only local structure is used during path generation. Wang et al. [38] in M-NMF algorithm, only use first order proximities along with community information. Thus, their method is not general enough to discover all the missing/future links such as long ranges, etc.

Based on figure 4, different proximities of the network along with community structure of the graph are important throughout the structural feature vector learning for the link prediction. We also compare our algorithm with some local neighbor based algorithms for the link prediction in figure 5. In this figure, we only use the structural feature vectors that are generated by DeepLink because the local neighborhood based methods only rely on the first order proximity.

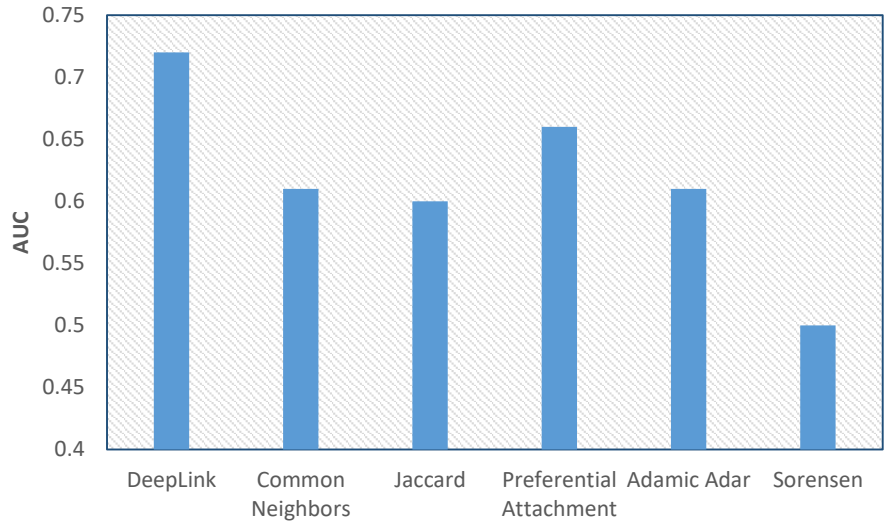

*Figure 5- Comparison of DeepLink against CN-based methods*

As indicated in figure 5, better structural features are extracted by DeepLink for link prediction. The local neighborhood based algorithms only focus on the first order proximities. While we consider higher order proximities alongside the community structure in DeepLink framework.

The detail examination of figures 4 and 5 also illustrates that considering the structural information along with the content information of the network nodes are efficient for the link prediction.

We also evaluate DeepLink on irBlogs dataset. Figure 6 depicts the effect of different representation learning algorithms on link prediction.

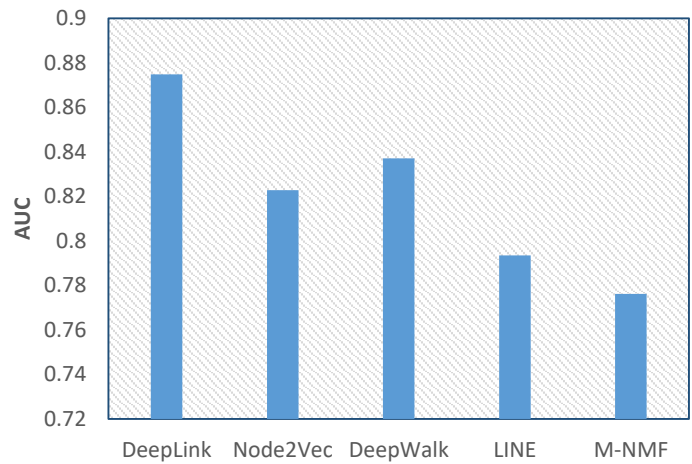

*Figure 6- AUC score for different network representation learning algorithms on irBlogs*

As can be seen in figure 6, DeepLink framework has a certain improvement compared with other network representation learning algorithms on irBlogs dataset. Similar to Telegram dataset, DeepLink outperforms the other methods because all the proximity orders and global structures along with content features of the network nodes are employed during feature learning. LINE and Node2vec show weak performance in comparison to Deepwalk due to ignoring higher order proximities. M-NMF has the worst results because the first order proximities are used in the algorithm. In addition, community and first-order proximity vectors are learned independently, then two vectors are combined to obtain final structural feature vector. So, it doesn't detect the missing/future links between the nodes with the distance of more than two. Figure 7 illustrates the comparison between DeepLink structural feature vectors and the CN-based algorithms, on irBlogs dataset.

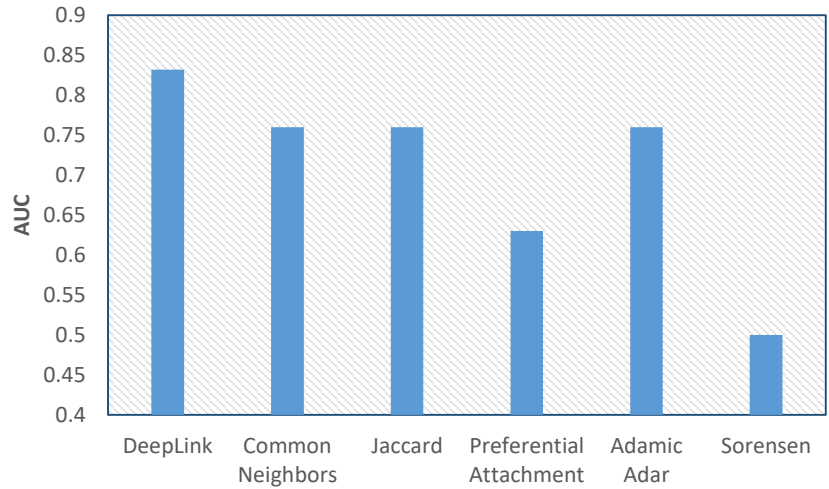

*Figure 7- Comparison of DeepLink against CN-based methods*

Both figures 5 and 7 indicate that the behavior of local structural based algorithms are different. This implies that these methods have different performance on different networks. This is due to the fact that in the local neighborhood based methods, only the first order proximities are considered. While in DeepLink framework, both local and community structural information for each node is employed to extract structural feature vectors. We also found that with the increase of geodesic distance, the performance of local neighborhood based methods gradually declines due to the loss of available network topology information. On the contrary, DeepLink framework performs consistently well, specifically when the geodesic distance is greater than two. This implies the learned feature vectors are comprehensive even when limited local structural information is available.

### 5.5. parameter sensitivity:

As it is previously stated, we concatenate structural feature vector with the content feature vector to generate the final feature vector for link prediction. The effect of different structural and content feature vectors dimension on Telegram dataset in DeepLink framework is investigated in figures 8 and 9. In these figures, the size of one of the feature vectors is considered constant. Then, the effects of different sizes for the other feature vector are examined in DeepLink.

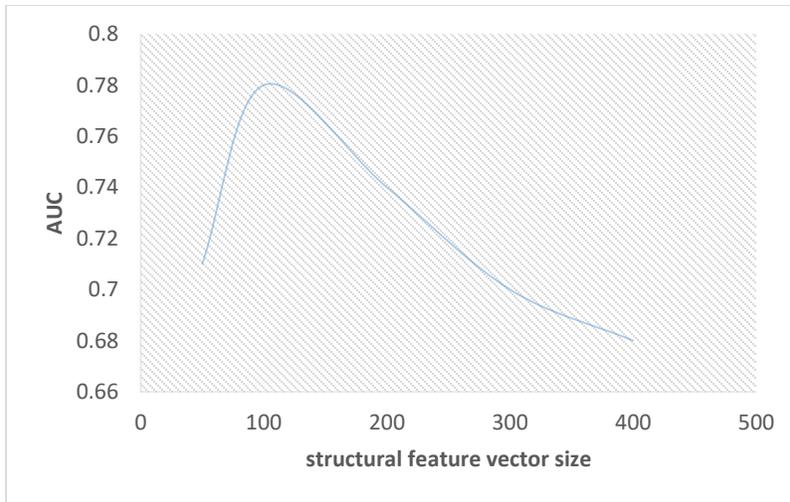

*Figure 8- The effects of structural feature vector size on Telegram*

In figure 8, the size of the content feature vector is considered 100. Then, the effect of vector size for structural features is investigated. As it is shown in figure 8, the best size of the structural feature vector is 100. When we choose a feature vector with the size of more than 100, irrelevant structural features are learned by DeepLink. In the other words, custom paths of each node which are generated by DeepLink enclose a set of nodes that are not related to the source node of the paths. Subsequently, unrelated information is embedded to Word2Vec model, thus its performance decreases rapidly. Also if the less contextual information is used into the framework, a number of existing links in the network would not be predicted. In Word2Vec model, the experimental results also indicate the best feature vector size is about 100. In figure 9, we show that the best size of the content feature vector is also 100.

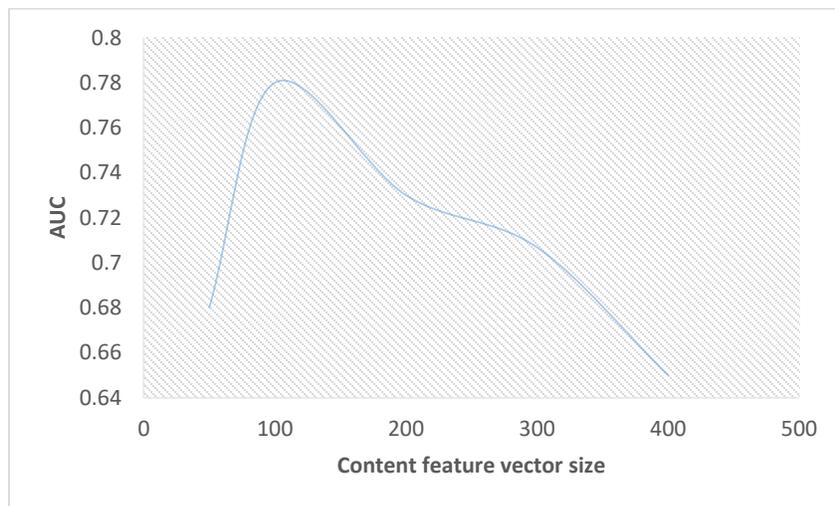

*Figure 9- The effects of content feature vector size on Telegram*

As can be seen in figures 8 and 9, in contrast to the most previous link prediction methods, both the structural and content feature vectors are equally important for link prediction. Furthermore, quality of contextual information which is generated by DeepLink is important. If unnecessary structural and content information are used for feature learning, the overall performance of the framework will be decreased.

## 6. Conclusion:

In this paper, we present a deep learning based framework for the link prediction. Many researchers have presented to predict new relationships in the network based on the structural information. Most of the features which are used in the previous researches, only consider local neighborhood structure such as common neighbors and number of the path between two nodes. While the recent studies show the impact of content features in the link prediction. In the proposed framework, both the structural and content features are used for the link prediction. The structural and content feature vectors are learned by Deep learning techniques. So, the features are extracted automatically. Unlike the previous researches, in the structural feature learning, we use all the structural information, including first and higher order proximities alongside the community structure of the network. The content feature vector is also contributed in DeepLink to find missing/future links in the network. The experimental results on two real social network datasets imply that both structural and content information are equally important for the link prediction. Besides, the features that are extracted by DeepLink framework are independent of the input networks while most of the previous studies cannot present the same behavior on different types of networks with various network topology.

In the future, we will consider the effect of different neural networks for feature learning on the link prediction. We will also investigate the effects of different topics of the nodes' contents on feature learning and link prediction.